\documentclass[12pt,a4paper]{article}
\usepackage[utf8]{inputenc}
\usepackage{amsmath}
\usepackage{amsfonts}
\usepackage{amssymb}
\usepackage{graphicx}
\usepackage{natbib}

\title{\bf Detection of the Binarity of the Star J1158+4239}

\author{M.Yu. Khovritchev$^{1}$\footnote{e-mail: deimos@gao.spb.ru}, 
A.M. Kulikova$^1$, E.N. Sokov$^{1,\,2}$, \\V.V. Dyachenko$^2$, 
D.A. Rastegaev$^2$, A.S. Beskakotov$^2$, 
\\Yu.Yu. Balega$^2$, B.S. Safonov$^3$, 
\\A.V. Dodin$^3$ and O.V. Vozyakova$^3$ \\
{\small $^1$Pulkovo Astronomical Observatory, Russian Academy of Sciences},\\
{\small Pulkovskoe sh. 65, St. Petersburg, 196140 Russia}\\
{\small $^2$Special Astrophysical Observatory, Russian Academy of Sciences,}\\
{\small Nizhnii Arkhyz, Karachai-Cherkessian Republic, 369167 Russia}\\
{\small $^3$Sternberg Astronomical Institute, Moscow State University,}\\
{\small Universitetskii pr. 13, Moscow, 119992 Russia}\\}

\begin{document}
\maketitle

\begin{abstract}

One of the goals of the Pulkovo program of research on stars with large proper motions is to reveal among the low-luminosity stars those that have evidence of binarity. Twelve astrometric binary candidates from the Pulkovo list have been included in the program of speckle observations with the BTA telescope at the Special Astrophysical Observatory of the Russian Academy of Sciences (SAO RAS) and the 2.5-m telescope at the Caucasus Mountain Observatory (CMO) of the Sternberg Astronomical Institute of the Moscow State University to confirm their binarity and then to determine the parameters of the revealed stellar pairs. The binarity of the brightest of these stars, J1158+4239 (GJ 3697), has been confirmed. Four sessions of speckle observations with the BTA SAO RAS telescope and one session with the 2.5-m CMO telescope have been carried out in 2015–2016. The weighted mean estimates of the pair parameters are $\rho$=286.5$\pm$1.2 mas and $\theta$=230.24$\pm$0.16$^{\circ}$ at the epoch B2015.88248. The magnitude difference between the pair stars is $\Delta m$=0.55$\pm$0.03 (a filter with a central wavelength of 800 nm and a FWHM of 100 nm) and $\Delta m$=0.9$\pm$0.1 (an R filter).
\\
{\bf DOI:} 10.1134/S1063773716100054
\end{abstract}
\section*{Introduction}
Low-mass dwarfs (M $\textless$ 0.6 M$_\odot$) account for almost three quarters of the Galactic stellar population, belong to various subsystems, and exhibit a considerable variety in the context of kinematics and physical parameters. Evidence of magnetic activity has been found in the outer shells of late-type dwarfs: flare and spot activity. These factors have determined a considerable interest in investigating low-mass stars in the last decade.

The luminosities of the stars under consideration are generally lower than the solar one by hundreds or thousands of times. Therefore, our knowledge about these objects is based largely on the results of observations of dwarfs in the immediate solar neighborhood (D $\textless$ 50 pc).

Attempts to construct and interpret the mass function near the “stars–brown dwarfs” boundary (see, e.g., \citet{Thies2015}) have shown the necessity of refining the fraction of binary and multiple objects among the dwarfs of various subsystems. When constructing present-day models of dwarf stars, researchers pay attention to the correspondence between the theoretical and observed mass–radius and mass–luminosity relations (\citet{Torres2013}; \citet{Spada2013}). Therefore, knowing accurate masses of dwarf stars plays an important role in testing such models.

These problems motivate the observers to search for binary systems among the dwarf stars, to construct their orbits, and to determine their masses (see, e.g., \citet{Cortes2015}; \citet{Opitz2016}). Positional observations of the motions of nearby dwarfs may play a significant role in these studies, because they allow astrometric binary stars within several tens of parsecs from the Sun to be revealed.

Astrometric binaries are separated into a special group by the detection technique. Usually we are dealing with a binary system for which the wobble of one of the components due to gravitational perturbations from the companion can be detected on the basis of astrometric observations. The second component is unobservable, for example, because of its very low luminosity. Occasionally, such objects are called stars with invisible companions. At present, the stars whose images are not resolved into components, but periodic deviations in the motion of their photocenter are noticeable are often attributed to astrometric binaries.

As a rule, solar-neighborhood stars are distinguished by large proper motions ($\mu\textgreater$ 0.1 arcsec$\cdot$yr$^{-1}$). Such stars are being actively investigated within the Pulkovo observational program (\citet{Khrutskaya2011}; \citet{Khovritchev2013}). Some of the program stars have turned out to exhibit evidence for nonlinearity of their motion over the celestial sphere (astrometric binary candidates; see \citet{Khovritchev2015}).

The most interesting objects have been selected to be tested for binarity through speckle observations with the BTA SAO RAS telescope, with which binary stars are being continuously observed and great experience has been accumulated (\citet{Balega2013}). This list includes a total of 12 stars. At present, the image of one of them, J1158+4239 (GJ 3697 or 2MASS 11585947+4239395, RA = 11$^h$ 58$^m$ 59.93$^s$, Dec = +42$^{\circ}$ 39$'$ 38.1$''$ at the epoch J2000), has been confidently separated into components, and the astrometric parameters of the pair and the magnitude difference between its components have been estimated. Additional observations of J1158+4239 have been performed with a speckle polarimeter at the Caucasus Mountain Observatory (CMO) of the Sternberg Astronomical Institute of the Moscow State University (SAI MSU). The details of these studies are reflected in the subsequent sections of this paper. The study of the remaining program stars continues.
\section*{Searching for astrometric binary stars}
\subsection*{\it Revealing the Nonlinearity of the Binary Photocenter Motion Based on Analysis of the Proper Motions}

For many of the binary systems consisting of low-mass stars, the angular separation between the components can be of the order of tenths of an arcsecond. Such binaries appear as single objects during observations with the instruments used for photographic and CCD sky surveys in the last decades. By measuring the coordinate of such stars based on the scans of photographic plates or CCD frames, we actually determine the position of the binary photocenter for the spectral band specified by the filter being used.

Given a dense and long series of observations, a periodicity of the motion can be revealed by standard methods. In most cases, we do not have such series for nearby dwarfs. Nevertheless, given several observations distributed nonuniformly in time, we can attempt to make estimates based on the proper motions. It is easy to see that the mean proper motion of the photocenter can differ significantly from the instantaneous one. A statistically significant difference between these proper motions compared to the errors of their determination can be revealed under certain conditions. This may be considered as evidence that the star in question is a binary system.

This idea was applied by \citet{Wielen1999} to bright stars from the FK6 catalog. When implementing the Pulkovo program of research on stars with large proper motions, this technique was adapted to reveal astrometric binaries among the low-luminosity stars in the solar neighborhood (\citet{Khrutskaya2011}; \citet{Khovritchev2015}).

The quantity F defined as $F^{2}=(\dfrac{\Delta\mu_{\alpha}cos\delta}{\epsilon_{\mu_{\alpha}}})^{2}+(\dfrac{\Delta\mu_{\delta}}{\epsilon_{\mu_{\delta}}})^{2}$ allows the significance of the difference $\mu_{inst}-\mu_{mean}$ to be estimated (here, $\Delta\mu$ are the corresponding proper motion differences, $\epsilon_{\mu}$ are the corresponding proper motion errors). At $F\textgreater2.49$ a star may be considered an astrometric binary candidate (or a $\Delta\mu$-binary candidate in Wielen’s terminology).

\subsection*{\it Implementing the Method Based on Sky Survey Data and Pulkovo Observations}

The present-day astrometric catalogs of stars constructed from CCD observations contain the coordinates for tens or hundreds of millions of stars with an accuracy of 20$-$100 mas, depending on the magnitude. Therefore, we decided to determine the proper motions of fast stars based on data from the CMC14, 2MASS, and SDSS catalogs. However, the epoch difference for these catalogs is typically 2$-$3 years, which is responsible for the low accuracy of the proper motions. Therefore, we performed CCD observations of more than 400 stars with $\mu\textgreater$ 0.3$''\cdot$yr$^{-1}$) at the Pulkovo normal astrograph (D = 330 mm, F = 3500 mm). As a result, we determined the equatorial coordinates and proper motions of these stars at an epoch difference of 5$-$10 years with an accuracy of 4$-$5 mas$\cdot$yr$^{-1}$ in the declination zone 30$^{\circ}-$70$^{\circ}$ .

The derived proper motions may be considered with some confidence as “quasi-instantaneous”. The proper motions presented in the LSPM catalog were determined at an epoch difference of 40$-$50 years and have the status of “quasi-mean” ones. The difference between the quasi-instantaneous and quasi-mean proper motions turned out to be significant for 70 stars. The data on this were published in \citet{Khrutskaya2011} and are accessible in electronic form in the Pulkovo astrometric database system~\footnote{http://puldb.ru} .

The described method of solving the problem has a number of problems. The stellar coordinates used in deriving the proper motions can be distorted by systematic errors, which are very difficult to reveal and to take into account. Therefore, we decided to make an attempt at studying the motions of fast stars by analyzing the images of sky fields from the DSS, 2MASS, SDSS DR12, and WISE digital surveys. The systematic errors of the stellar coordinates were investigated for each individual survey. To calculate the proper motions, we formed a unified system of reference stars (i.e., the reference stars were the same in all frames and plate scans). As a result of these studies, the proper motions were determined for 1308 stars in the declination zone 30$^{\circ}-$70$^{\circ}$ with a mean accuracy of 4 mas$\cdot$yr$^{-1}$. Evidence for nonlinearity of the motion over the celestial sphere is observed for 121 stars from this list (\citet{Khovritchev2015}). All the results of this study are accessible in the VizieR~\footnote{http://cdsarc.u-strasbg.fr/viz-bin/Cat?J/PAZh/41/896} database.

\subsection*{\it Motion of the Photocenter of J1158+4239}

The investigated objects should be qualified as astrometric binary candidates. Further studies are required to confirm their binarity. For 12 stars from the Pulkovo program characterized by large F , we proposed to perform speckle observations with the BTA SAO RAS and CMO SAI MSU telescopes.

The star J1158+4239 with a V magnitude of 12.09$^m$ was one of the brightest in the list and, therefore, most promising. Its proper motion components in the LSPM catalog are -330 and 61 mas$\cdot$yr$^{-1}$ in RA and Dec, respectively. Its quasi-instantaneous proper motion in Khrutskaya et al. (2011) turned out to be -303.3$\pm$8.8 and 114.1$\pm$9.0 mas$\cdot$yr$^{-1}$ , which provided F = 4.44.

\begin{figure}[h!]
\center{\includegraphics[width=1\linewidth]{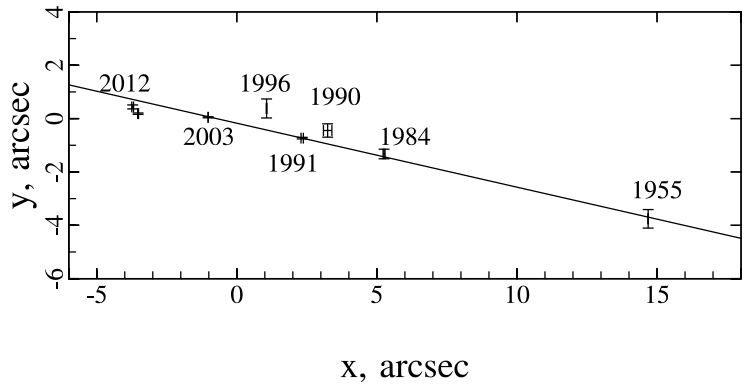}}
\caption{\small Motion of the photocenter of J1158+4239 from photographic and digital sky surveys (1955 -- POSS1, 1984 -- GSC1, 1990 and 1991 -- POSS2, 1996 -- GSC2, 2003 -- SDSS, 2012 -- Pulkovo observations). The line shows the mean motion of the photocenter from all of the observations used; x and y are the tangential coordinates (the projection center is specified by the equatorial coordinates at the mean epoch of observations, the x and y axes are directed to the east and the north, respectively).}
\label{fig1}
\end{figure}

Figure 1 shows the motion of the photocenter for J1158+4239. For all of the surveys used (1955 -- POSS1, 1984 -- GSC1, 1990 and 1991 -- POSS2, 1996 -- GSC2, 2003 -- SDSS, 2012 -- Pulkovo observations), we managed to select eight reference stars whose images are reliably fitted in all frames with an acceptable signal-to-noise ratio. As a result, for the mean motion of the photocenter over all points the proper motion components turned out to be $-$326.4 $\pm$ 1.7 and 50.1 $\pm$ 7.1 mas$\cdot$yr$^{-1}$. For the quasi-mean proper motion (1955 -- 1996) we obtained $-$335.9 $\pm$ 6.4 and 84.7 $\pm$ 6.3 mas$\cdot$yr$^{-1}$. The quasi-instantaneous proper motion (2003 -- 2012) was $-$303.7 $\pm$ 8.5 and 21.1 $\pm$ 7.9 mas$\cdot$yr$^{-1}$. The value of the test F = 6.99, which is consistent with its previous estimates (the star may be considered as an astrometric binary candidate).

\section*{Speckle observations}
\subsection*{\it Observations with the BTA SAO RAS Speckle Interferometer}

Our speckle observations were carried out with the 6-m SAO RAS telescope in June, October, December 2015, and February 2016. We used a speckle interferometer based on an Andor iXon Ultra 512$\times$512-pixel EMCCD camera. The field size was 4.5$\times$4.5 arcsec. The scales were calibrated by recording the interference pattern from a two-hole mask located on the covers of the telescope’s prime-focus cabin. The position was calibrated by measuring the position angles of the Orion Trapezium components ($\theta^1$ Ori). We used an interference filter with a central wavelength of 800 nm and a FWHM of 100 nm. 

\begin{figure}[h]
\begin{minipage}[h]{0.38\linewidth}
\center{\includegraphics[width=0.99\linewidth]{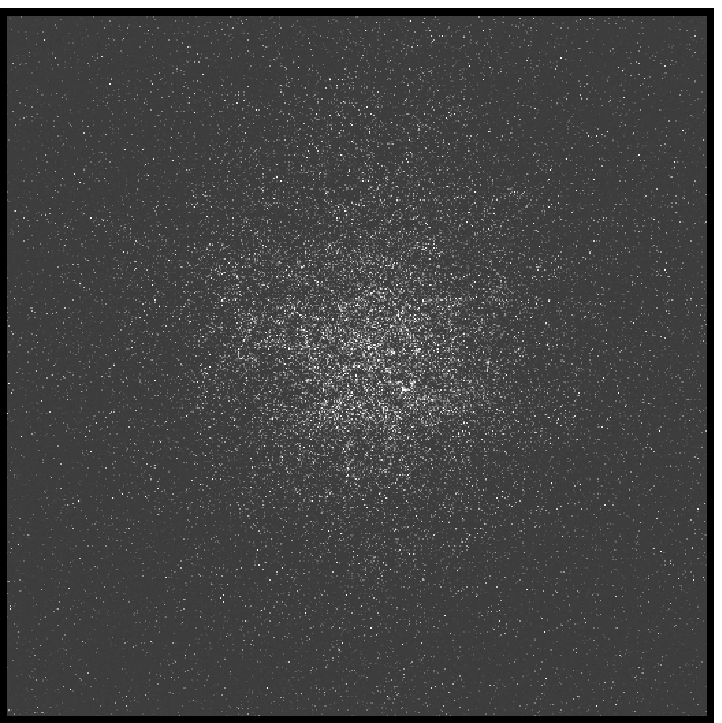}}
\end{minipage}
\hfill
\begin{minipage}[h]{0.62\linewidth}
\center{\includegraphics[width=0.99\linewidth]{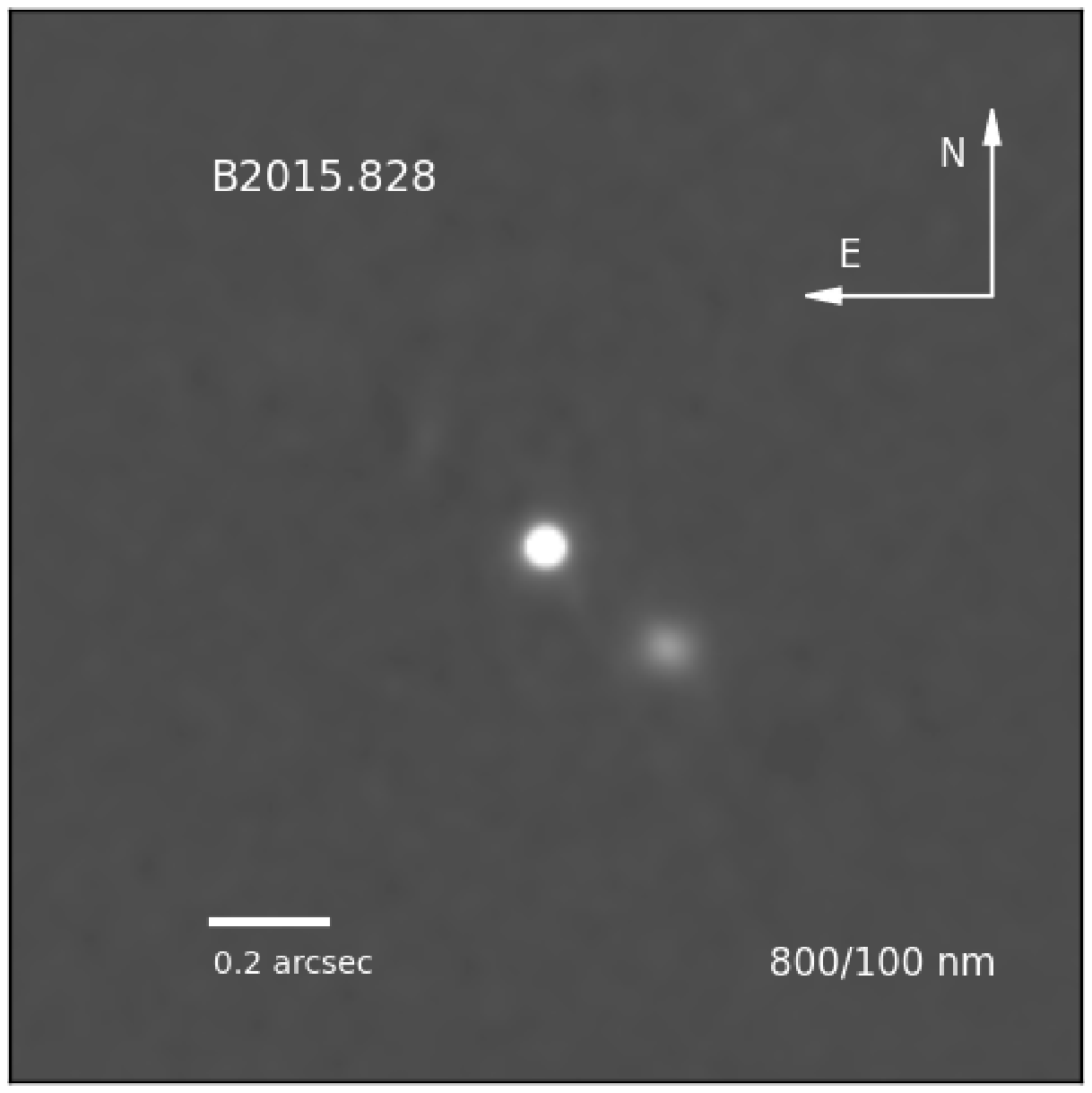}}
\end{minipage}
\caption{\small \emph{Left:} BTA SAO RAS speckle interferogram of J1158+4239. \emph{Right:} The corresponding reconstructed image of the binary system J1158+4239.}
\label{ris:fig2}
\end{figure}

During the observations we recorded several series of 2000 short-exposure speckle images. The exposure time was 20 ms, the total series accumulation time, including the expenditure on the frame reading and recording, was about 3 min. The data were reduced by a standard method (Balega et al. 2002; Pluzhnik 2005). The power spectrum was obtained from each frame, whereupon it was averaged for the entire series. The positional parameters and the magnitude difference (for the final results, see the table) were obtained from the averaged power spectrum through simulations. The ambiguity of 180 degrees was removed by calculating the phase by the method of bispectral analysis (Lohmann et al. 1983). Examples of the speckle interferogram and the re constructed image for J1158+4239 are presented in Fig. 2.

\subsection*{\it Speckle Observations at the Caucasus Mountain Observatory of the SAI MSU}

The observations of J1158+4239 were also performed with the speckle polarimeter (\citet{Safonov2015})~\footnote{http//lnfm1.sai.msu.ru/kgo/instruments/mfc/SPPdesign.pdf} of the 2.5-m CMO SAI MSU telescope. In its design, the instrument is a camera in which two images corresponding to the horizontal and vertical polarizations are formed. The effective angular scale of the camera is small enough to provide a proper calculation of the object’s spatial spectrum up to the cutoff frequency. The detector is an Andor iXon+897 CCD with electronic amplification.

With this instrument we obtained a series with a length of $\approx$9500 frames at a single-frame exposure time of 30 ms, the mid-series time is 23:03:49 UT on January 15, 2016. The filter was a standard Bessel R filter. Note that the transmission curve of this filter has a fairly large FWHM, which leads to an isotropic drop in the contrast of the speckle pattern. The image blurring due to atmospheric dispersion and beamsplitter dispersion also begins to play a significant role. Both these effects lower the signal-to-noise ratio in the power spectrum but not fatally, and they were taken into account in the reduction.

\begin{figure}[h!]
\center{\includegraphics[width=1\linewidth]{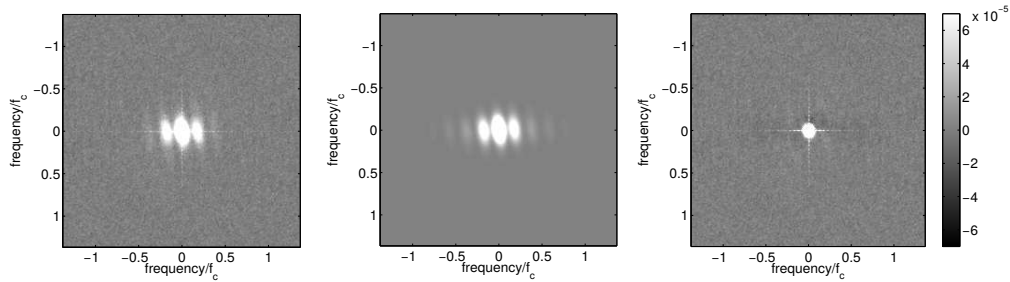}}
\caption{\small \emph{Left:} Averaged power spectrum for the entire series, the right polarimeter beam. \emph{Central:} Simulated power spectrum, the source’s parameters are given in the text. \emph{Right:} Observation–model difference. The spatial frequency normalized to the cutoff frequency of the optical system is along the axes. The brightness corresponds to the logarithm of the power spectrum, the scale is given on the right.}
\label{ris:fig3}
\end{figure}

To estimate the binary parameters, we applied the standard approach of speckle interferometry, i.e., fitting the averaged power spectrum (\citet{Tokovinin2010}). A binary source was taken as a model of the object. The fitting was done by minimizing the total residual by the Levenberg–Markquardt method. The averaged power spectrum and its model are shown in Fig. 3. The fringes are evidence of the source’s binarity.

To estimate the errors in the binary parameters, we divided the entire series into three equal parts and considered separately the images referring to two polarimeter beams. The parameters were sought separately for each of the six sets; we provide the means of their values as the final result and the rms deviations are the estimates of their errors. The binary parameters in the detector’s reference frame are as follows: the separation $\rho$ = 14.37 $\pm$ 0.07 pixels, the position angle $\theta$ = 172.8$^{\circ}$ $\pm$ 0.4$^{\circ}$ , and the contrast $\Delta m$ = 0.9$^m$ $\pm$ 0.1$^m$ . We removed the uncertainty in the position angle of 180$^{\circ}$ , which inevitably arises in this method of estimating the binary parameters, by comparing the profile of the averaged bispectrum of the frames and the calculated bispectrum for two models with $\theta$ = 172.8$^{\circ}$ and $-$7.2$^{\circ}$ . In the former case, the agreement is much better.

By observing three wide pairs with the speckle interferometer and an ordinary CCD with a large field of view, we estimated the angular scale of the camera, 20.19 $\pm$ 0.16 mas/pixel, and the position-angle correction, 0.9 $\pm$ 0.4$^{\circ}$ (for more details, see \citet{Safonov2016})~\footnote{http://lnfm1.sai.msu.ru/kgo/instruments/mfc/SPPplatepar.pdf} . Using these values and Eq. (2) from \citet{Safonov2016}, we converted the binary parameters to the equatorial coordinate system: the separation $\rho$ = 295 $\pm$ 3 mas and the position angle $\theta$ = 229.6$^{\circ}$ $\pm$ 0.7$^{\circ}$ .

\begin{table}[!ht]
\caption{Speckle measurements of J1158 + 4239.}
\smallskip
\begin{center}
{\small
\begin{tabular}{l|l|ll|ll|ll}
\hline
\noalign{\smallskip}
Epoch      & Observatory & $\Delta m$~~~~~~ & $\sigma _{\Delta m}$ & $\rho, mas$ ~~~~& $\sigma _\rho$ & $\theta, \deg $~~~~~~& $\sigma_\theta$ \\
\noalign{\smallskip}
\hline
\noalign{\smallskip}
2015.43317 & SAO RAS        & 0.48       & 0.04     & 282     & 3  & 228.6 & 0.5               \\
2015.82798 & SAO RAS       & 0.54       & 0.03     & 284     & 1  & 230.2 & 0.1               \\
2015.97313 & SAO RAS       & 0.59       & 0.03     & 287     & 1  & 230.2 & 0.1               \\
2016.04122 & SMO SAU MSU   & 0.90       & 0.10     & 295     & 3  & 229.6 & 0.7               \\
2016.13692 & SAO RAS        & 0.70       & 0.08     & 288     & 1  & 230.9 & 0.2               \\
\noalign{\smallskip}
\hline
\end{tabular}
}
\end{center}
{\small For the SAO RAS and CMO SAI MSU observations we used an interference filter with a central wavelength of 800 nm and a FWHM of 100 nm and a standard R filter, respectively. The binary parameters:  $\Delta m$ is the contrast, $\rho$ is the separation, $\theta$ is the position angle. $\epsilon_{\Delta m}$, $\epsilon_{\theta}$ and $\epsilon_{\rho}$ are the corresponding parameter errors.}
\end{table}

\section*{Discussion}

The binary J1158+4239 is absent in the WDS catalog and any other lists of binary stars. Thus, it can be said that its binarity has been detected for the first time. It is to be studied further to determine the orbit and the component masses. 

The weighted mean estimates of the observed parameters for J1158+4239 at the epoch B2015.88248 are $\rho$ = 286.5 $\pm$ 1.2 mas and $\theta$ = 230.24 $\pm$ 0.16$^{\circ}$ . The magnitude difference between the components obtained with the BTA SAO RAS telescope is $\Delta m$ = 0.55 $\pm$ 0.03 (a filter with a central wavelength of 800 nm and a FWHM of 100 nm). The CMO observations give $\Delta m$ = 0.9 $\pm$ 0.1 (an R filter). The astrometric parameters determined with the BTA SAO RAS and 2.5-m CMO telescopes are in satisfactory agreement between themselves (overlap within 2$\sigma$).

At present, the trigonometric parallax is unknown for J1158+4239. The photometric and spectroscopic parallax estimates are presented in \citet{Lepine2013} and are 39 $\pm$ 10 and 38 $\pm$ 11 mas, respectively. This corresponds to a distance of $\approx$25 pc. Thus, the components of J1158+4239 are dwarfs located in the immediate solar neighborhood. The orbital period in this binary system was roughly estimated from the available data (parallax and angular separation) to be about 20 years. In the same paper there is an indication that the spectral type is M2.0. \citet{Pickles2010} determined the spectral type as M1.9V.

Photometric data from various instruments are available for J1158+4239. SDSS (\citet{Alam2015}) and 2MASS (\citet{Cutri2003}) are known for the high accuracy of the magnitude estimates in the corresponding bands. The magnitude estimates for SDSS are $u$ = 16.140$^m$ , $g$ = 13.083$^m$ , $r$ = 11.530$^m$, $i$ = 10.534$^m$ , and $z$ = 10.317$^m$ with $\pm$0.001$^m$ errors (except the $u$ band; the error here is $\pm$0.01$^m$ ). In 2MASS this star is characterized by the magnitude estimates $J$ = 8.638$^m$ $\pm$ 0.023$^m$ , $H$ = 8.047$^m$ $\pm$ 0.026$^m$ , and $Ks$ = 7.825$^m$ $\pm$ 0.015$^m$ .

These data allow us to estimate which components can form the corresponding colors. As a tool for our calculations we used the service~\footnote{http://stev.oapd.inaf.it/cgi-bin/cmd} that allows the curves of normal colors to be constructed from the Padova isochrones (\citet{Bressan2012}). Thus, we can get the tables containing the masses and the absolute magnitudes in SDSS and 2MASS bands for stars with given metallicities and ages. By combining stars with different masses, it is easy to select the pairs giving colors from $u-g$ to $H-Ks$ so that their difference from the real SDSS and 2MASS data was minimal. For J1158+4239 such a selection gives component masses of 0.42 M$_{\odot}$ and 0.36 M$_{\odot}$.

\begin{figure}[h!]
\center{\includegraphics[width=0.7\linewidth]{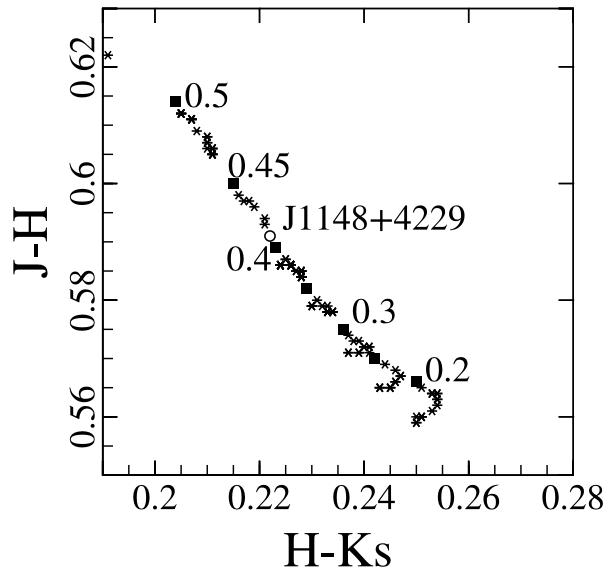}}
\caption{\small The curve of normal colors for low-mass main-sequence dwarfs in ($H-Ks$)$-$($J-H$) coordinates constructed from the Padova isochrones. The squares correspond to stars with masses from 0.2$M_\odot$ to 0.5$M_\odot$ with a step of 0.05$M_\odot$. The asterisks indicate the locations of the pairs formed by a star with the current mass (squares) and a star with a lower mass in the range from the current mass to 0.09$M_\odot$. The position of J1158+4239 is indicated by the open circle.}
\label{ris:fig4}
\end{figure}

Figure 4 shows the curve of normal colors in ($H-Ks$)$-$($J-H$) coordinates. The squares correspond to stars with masses from 0.2 M$_{\odot}$ to 0.5 M$_{\odot}$ with a step of 0.05 M$_{\odot}$. The asterisks indicate the locations of the pairs formed by a star with the current mass (squares) and a star with a lower mass in the range from the current mass to the minimum one (for the Padova isochrones at present this is 0.09 M$_{\odot}$). Thus, we see how the colors are formed when components of different masses are mixed. The position of J1158+4239 is indicated by the open circle. The difference of the model and real colors in ($H-Ks$)$-$($J-H$) coordinates is 0.002$^m$ (i.e., the model point almost coincides with the real one on the scale of Fig. 4).

One cannot say that the models of low-mass stars are characterized by a high accuracy (Spada et al. 2013). This may explain the fact that the agreement for other colors is poorer than that for the infrared data and is hundredths (and occasionally tenths) of a magnitude.

The magnitude differences obtained during speckle observations provide an additional possibility for testing these constructions. The SDSS i filter corresponds to the interference filter used in the BTA SAO RAS observations better than other filters of this survey. The model stellar pair for this filter gives a magnitude difference of 0.5 m . The model magnitude difference for the R band is 0.58 m . It can be said with some caution that the BTA SAO RAS magnitude estimates agree with the results of our model selection of components. Only one observation has been performed to date at the CMO SAI MSU; therefore, it is still too early to say that the estimates of the component magnitude difference in the R band differ significantly from the model selected on the basis of Padova isochrones.

\section*{Acknowledgments}

This work was supported in part by Program P-41 of the Presidium of the Russian Academy of Sciences (“Transitional and Explosive Processes in Astrophysics”). The production of the speckle polarimeter for the 2.5-m telescope is supported by the MSU Development Program and the Russian Foundation for Basic Research (project no. 16-32-60065). E.N. Sokov thanks the Russian Science Foundation (project no. 14-50-00043, the area of research—Exoplanets) for its support in performing and analyzing the speckle observations at the 6-m SAO RAS telescope. This study would not be possible without astronomical data from the VizieR (CDS, Strasbourg, France), SDSS (Sloan Digital Sky Survey), STScI MAST (Mikulski Archive for Space Telescopes), and IRSA (Infrared Science Archive) databases. We are grateful to the developers of these resources. We thank the referees for a careful reading of the original text and for the remarks that allowed the presentation of the material to be improved.

\end{document}